\documentclass[12pt]{article}
\usepackage{amsmath}
\usepackage{amssymb}
\usepackage{graphicx,psfrag,epsf}
\usepackage{enumerate}
\usepackage[numbers]{natbib}
\usepackage{bm}
\usepackage{hyperref}

\newcommand{\blind}{0}

\setcitestyle{square}

\usepackage[]{xcolor}
\makeatletter
\def\maxwidth{ %
  \ifdim\Gin@nat@width>\linewidth
    \linewidth
  \else
    \Gin@nat@width
  \fi
}
\makeatother

\definecolor{fgcolor}{rgb}{0.345, 0.345, 0.345}
\newcommand{\hlnum}[1]{\textcolor[rgb]{0.686,0.059,0.569}{#1}}%
\newcommand{\hlstr}[1]{\textcolor[rgb]{0.192,0.494,0.8}{#1}}%
\newcommand{\hlcom}[1]{\textcolor[rgb]{0.678,0.584,0.686}{\textit{#1}}}%
\newcommand{\hlopt}[1]{\textcolor[rgb]{0,0,0}{#1}}%
\newcommand{\hlstd}[1]{\textcolor[rgb]{0.345,0.345,0.345}{#1}}%
\newcommand{\hlkwa}[1]{\textcolor[rgb]{0.161,0.373,0.58}{\textbf{#1}}}%
\newcommand{\hlkwb}[1]{\textcolor[rgb]{0.69,0.353,0.396}{#1}}%
\newcommand{\hlkwc}[1]{\textcolor[rgb]{0.333,0.667,0.333}{#1}}%
\newcommand{\hlkwd}[1]{\textcolor[rgb]{0.737,0.353,0.396}{\textbf{#1}}}%

\usepackage{framed}
\makeatletter
\newenvironment{kframe}{%
 \def\at@end@of@kframe{}%
 \ifinner\ifhmode%
  \def\at@end@of@kframe{\end{minipage}}%
  \begin{minipage}{\columnwidth}%
 \fi\fi%
 \def\FrameCommand##1{\hskip\@totalleftmargin \hskip-\fboxsep
 \colorbox{shadecolor}{##1}\hskip-\fboxsep
     \hskip-\linewidth \hskip-\@totalleftmargin \hskip\columnwidth}%
 \MakeFramed {\advance\hsize-\width
   \@totalleftmargin\z@ \linewidth\hsize
   \@setminipage}}%
 {\par\unskip\endMakeFramed%
 \at@end@of@kframe}
\makeatother

\definecolor{shadecolor}{rgb}{.97, .97, .97}
\definecolor{messagecolor}{rgb}{0, 0, 0}
\definecolor{warningcolor}{rgb}{1, 0, 1}
\definecolor{errorcolor}{rgb}{1, 0, 0}
\newenvironment{knitrout}{}{} 

\usepackage{alltt}
\IfFileExists{upquote.sty}{\usepackage{upquote}}{}

\begin{document}

\def\spacingset#1{\renewcommand{\baselinestretch}%
{#1}\small\normalsize} \spacingset{1}


\if0\blind
{
  \title{\bf mixdistreg: An R Package for Fitting\\ Mixture of Experts Distributional Regression\\ with Adaptive First-order Methods}
  \author{David R\"ugamer\\
    \small{Department of Statistics, TU Dortmund}\\
    \small{Department of Statistics, LMU Munich}\\
    \small{Munich Center for Machine Learning}}
  \maketitle
} \fi

\bigskip

\begin{abstract}
    This paper presents a high-level description of the R software package \texttt{mixdistreg} to fit mixture of experts distributional regression models. The proposed framework is implemented in R using the deepregression software template, which is based on TensorFlow and follows the neural structured additive learning principle. The software comprises various approaches as special cases, including mixture density networks and mixture regression approaches. Various code examples are given to demonstrate the package's functionality.
\end{abstract}

\section{Summary}
\label{sec:summary}

Mixture models are a common choice when data stems from different sub-populations, but only the pooled data with unknown membership is observed. Models are typically estimated using the EM algorithm. As the log-likelihood of mixture models is, in general, not convex \cite{Murphy.2012}, existing optimization techniques are vulnerable to ending up in local optima \cite{Chaganty.2013}. In contrast, mixture density networks \cite[MDN;][]{Bishop.1994} that can also be considered as a type of mixture regression models, are successfully optimized with first-order (gradient descent) methods. \cite{nmdr} therefore proposed the mixture of experts distributional regression, a combination of interpretable mixtures of distributional regression models and MDNs to facilitate robust estimation and extend mixtures of regression models to the distribution regression case. The framework in \cite{nmdr} is based on a neural network formulation implemented in \texttt{mixdistreg} (\url{https://github.com/neural-structured-additive-learning/mixdistreg}), which is presented in this paper.

\section{Statement of need}
\label{sec:need}

Common EM optimization routines are limited in their flexibility to specify mixtures of (many) potentially different distributions, cannot cope with large amounts of data and, in particular, are not robust in high dimensions. First-order methods used in deep learning are applied on mini-batches of data allowing large data set applications and can be used in a generic fashion for all model classes. In order to obtain a scalable framework, we implement the proposed framework by \cite{nmdr} in R \cite{R} using the software template \texttt{deepregression} \cite{rugamer2021deepregression} which relies on TensorFlow \cite{TensorFlow}. The template described in \cite{rugamer2021deepregression} provides the basis for many other neural network-based modeling approaches referred to as \emph{neural structured additive learning (NSAL)}. Examples include neural-based and autoregressive transformation models \cite{Baumann.2020, deeptrafo2021, kook2020ordinal, dats2021}, survival regression in neural networks \cite{bender2020, kopper2020, kopper2021}, distributional regression \cite{rugamer2020unifying} or scalable factor models and factorizations \cite{fdtf2021, AFM}. The NSAL principle is also followed by \texttt{mixdistreg} which allows combining it with other neural-based approaches straightforwardly. The software further comprises many different other approaches as a special case, including neural density networks \cite{Magdon.1998}, MDNs, and various mixture regression approaches (with penalized smooth effects) as proposed in \cite{Leisch.2004, Gruen.2007, Stasinopoulos.2007}. 

\section{Implementation Details}
\label{sec:details}

In the following, we briefly describe the main function \texttt{mixdistreg} of the eponymous package.

Given a realization $y$ of the outcome of interest $Y$ and features $\bm{x} \in \mathbb{R}^p$, the package models the following density:
\begin{equation} \label{eq:model}
    f_{Y\mid\bm{x}}(y \mid \bm{x}) = \sum_{m=1}^M \pi_m (\bm{x}) f_m(y\mid\bm{\theta}_m(\bm{x})),
\end{equation}
where $f_m$ are density functions of (potentially different) distributions $\mathcal{F}_m, m=1,\ldots,M$ with parameters $\bm{\theta}_m = (\theta_{m,1},\ldots,\theta_{m,k_m})^\top$, and $\pi_m \in [0,1]$ are mixture weights that sum to 1. 

\paragraph{Key Features:}
\texttt{mixdistreg} allows, among other things, for
\begin{itemize}
    \item mixtures of the same parametric distributions ($\mathcal{F}_m \equiv \mathcal{F}^\ast \,\,\forall m\in\{1,\ldots,M\}$);
    \item mixtures of different parametric distributions with the same domain ($\mathcal{F}_m \neq \mathcal{F}_n$ for some $m,n\in\{1,\ldots,M\}, m\neq n$);
    \item mixture components ($f_m$) to be chosen from a variety of distributions \cite[see][]{rugamer2021deepregression};
    \item defining distribution parameters via an additive predictor $\eta$ and a link function $g$ (i.e., $g^{-1}(\theta(\bm{x})) = \eta(\bm{x})$ and additive structure $\eta(\bm{x}) = \sum_{j=1}^J \eta_j(\bm{x})$)
    \item individual additive predictors for different mixture components $m$ (i.e., $g^{-1}(\theta_{m}(\bm{x})) = \eta_{m}(\bm{x})$);
    \item individual additive predictors and link functions for different distribution parameters $k$ within one mixture component $m$ (i.e., $g^{-1}_{m,k}(\theta_{m,k}(\bm{x})) = \eta_{m,k}(\bm{x})$);
    \item a separate model definition that relates the categorical distribution to features of all sorts (i.e., $g^{-1}_\pi(\pi_{m}(\bm{x})) = \sum_{j=1}^{J_\pi} \eta_{m,j}(\bm{x})$);
    \item mixtures with one-point degenerate distributions (i.e., ($f_m(y) = 1_{y=a}, a\in\mathbb{R}$)  as, e.g., used in zero-inflated models \cite[see, e.g.,][]{fritz2021combining}.
\end{itemize}
A formula interface for the different (or same) mixture components follows the intuitive \texttt{S}-like formula interface to define the additive predictor functions $\eta$. These can be defined flexibly, including linear effects (e.g., $\eta_j(\bm{x}) = \bm{x}^\top \bm{\beta}$), smooth terms, $L_1$-penalized sparse effects, (deep) neural network components or a combination thereof \cite[for details see][]{rugamer2021deepregression}. 

\paragraph{Convenience Functions:}
Several convenience functions exist, wrapping this main function:
\begin{itemize}
    \item \texttt{sammer}: a simpler interface for \textbf{sam}e \textbf{m}ixtur\textbf{e} (distributional) regression
    \item \texttt{inflareg}: a simpler interface for \textbf{infl}ated \textbf{reg}ression models forming a mixture of one or more degenerate distribution(s) and another parametric distribution;
    \item \texttt{zinreg, oinreg, zoinreg}: convenience functions, in turn, wrapping \texttt{inflareg} to allow for a simple model definition of zero-inflated, one-inflated and zero-and-one-inflated regression models.
\end{itemize}
Next to these modeling functions, the package provides a function \texttt{gen\_mix\_dist\_maker} to allow for more complex user-defined mixtures (e.g, mixtures of various parametric distributions and one-point mass distributions), and methods for plotting (\texttt{plot}), obtaining model coefficients (\texttt{coef}), calculating posterior probabilities for all clusters (\texttt{get\_pis}) and extracting statistics of the mixture components (\texttt{get\_stats\_mixcomps}).

\section{Examples}
The following examples demonstrate the interface of \texttt{mixdistreg}.

\subsection{Mixture of Linear Regressions}

We start with a simple special case, the mixture of linear regressions, and compare the results with \texttt{flexmix} \citep{Leisch.2004}, a well-established package for mixtures of regression models in R.

\begin{figure}[!h]
    \centering
    \includegraphics[width=0.8\maxwidth]{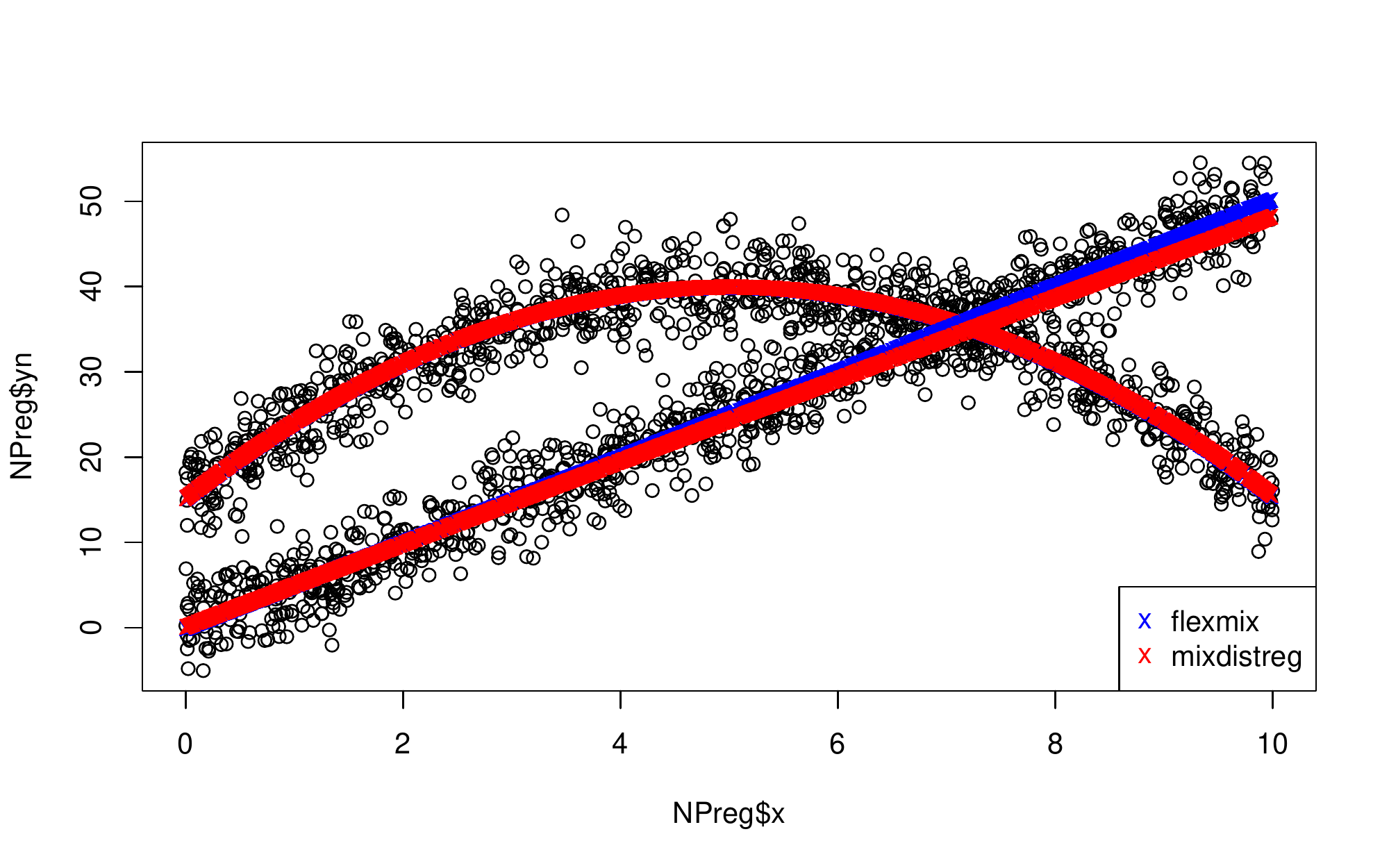}
    \caption{Comparison of \texttt{flexmix} (blue) and \texttt{mixdistreg} (red) results on the \texttt{NPreg} data showing a mixture of two trends in the variable \texttt{x}, which are found by both methods.}
    \label{fig:1}
\end{figure}

\begin{knitrout}
\definecolor{shadecolor}{rgb}{0.969, 0.969, 0.969}
\color{fgcolor}
\begin{kframe}
\begin{alltt}
\hlkwd{library}\hlstd{(flexmix)}
\hlkwd{library}\hlstd{(dplyr)}
\hlkwd{library}\hlstd{(mixdistreg)}
\end{alltt}

\begin{alltt}
\hlcom{# Load the data}
\hlkwd{set.seed}\hlstd{(}\hlnum{42}\hlstd{)}
\hlstd{NPreg} \hlkwb{<-} \hlkwd{ExNPreg}\hlstd{(}\hlkwc{n} \hlstd{=} \hlnum{1000}\hlstd{)}
\hlstd{nr_comps} \hlkwb{<-} \hlnum{2}
\hlstd{NPreg}\hlopt{$}\hlstd{xsq} \hlkwb{<-} \hlstd{NPreg}\hlopt{$}\hlstd{x}\hlopt{^}\hlnum{2}
\end{alltt}

\begin{alltt}
\hlcom{# Fit a mixture of regression models with mixtools}
\hlkwd{set.seed}\hlstd{(}\hlnum{42}\hlstd{)}
\hlstd{fm_mod} \hlkwb{<-} \hlkwd{flexmix}\hlstd{(yn} \hlopt{~} \hlstd{x} \hlopt{+} \hlstd{xsq,}
                  \hlkwc{data} \hlstd{= NPreg,}
                  \hlkwc{k} \hlstd{= nr_comps)}
\end{alltt}

\begin{alltt}
\hlcom{# Fitted values}
\hlstd{pred_fm} \hlkwb{<-} \hlstd{fm_mod} \hlopt{%>%} \hlkwd{predict}\hlstd{()}
\end{alltt}

\begin{alltt}
\hlcom{# Fit a mixture of normal regression with mixdistreg}
\hlstd{dr_mod} \hlkwb{<-} \hlkwd{sammer}\hlstd{(}\hlkwc{y} \hlstd{= NPreg}\hlopt{$}\hlstd{yn,}
                 \hlkwc{family} \hlstd{=} \hlstr{"normal"}\hlstd{,}
                 \hlkwc{nr_comps} \hlstd{= nr_comps,}
                 \hlkwc{list_of_formulas} \hlstd{=} \hlkwd{list}\hlstd{(}\hlkwc{mean} \hlstd{=} \hlopt{~} \hlnum{1} \hlopt{+} \hlstd{x} \hlopt{+} \hlstd{xsq,}
                                         \hlkwc{scale} \hlstd{=} \hlopt{~}\hlnum{1}\hlstd{),}
                 \hlkwc{data} \hlstd{= NPreg,}
                 \hlkwc{optimizer} \hlstd{=} \hlkwd{optimizer_rmsprop}\hlstd{(}\hlkwc{learning_rate} \hlstd{=} \hlnum{0.01}\hlstd{),}
                 \hlkwc{tf_seed} \hlstd{=} \hlnum{42}
\hlstd{)}
\end{alltt}

\begin{alltt}
\hlcom{# Train network}
\hlstd{dr_mod} \hlopt{%>%} \hlkwd{fit}\hlstd{(}\hlkwc{epochs} \hlstd{=} \hlnum{5000L}\hlstd{,}
               \hlkwc{validation_split} \hlstd{=} \hlnum{0.1}\hlstd{,}
               \hlkwc{patience} \hlstd{=} \hlnum{100L}\hlstd{,}
               \hlkwc{early_stopping} \hlstd{=} \hlnum{TRUE}\hlstd{,}
               \hlkwc{verbose} \hlstd{=} \hlnum{FALSE}\hlstd{)}
\end{alltt}

\begin{alltt}
\hlcom{# Fitted means of normal distributions}
\hlstd{pred_dr} \hlkwb{<-} \hlstd{dr_mod} \hlopt{%>%} \hlkwd{get_stats_mixcomps}\hlstd{(}\hlkwc{what} \hlstd{=} \hlstr{"means"}\hlstd{)}
\hlcom{# Compare}
\hlkwd{plot}\hlstd{(NPreg}\hlopt{$}\hlstd{yn} \hlopt{~} \hlstd{NPreg}\hlopt{$}\hlstd{x)}
\hlkwa{for}\hlstd{(i} \hlkwa{in} \hlnum{1}\hlopt{:}\hlstd{nr_comps)\{}
  \hlkwd{points}\hlstd{(pred_fm[[i]]} \hlopt{~} \hlstd{NPreg}\hlopt{$}\hlstd{x,} \hlkwc{col}\hlstd{=}\hlstr{"blue"}\hlstd{,} \hlkwc{pch}\hlstd{=}\hlstr{"x"}\hlstd{)}
  \hlkwd{points}\hlstd{(pred_dr[,i]} \hlopt{~} \hlstd{NPreg}\hlopt{$}\hlstd{x,} \hlkwc{col}\hlstd{=}\hlstr{"red"}\hlstd{,} \hlkwc{pch}\hlstd{=}\hlstr{"x"}\hlstd{)}
\hlstd{\}}
\hlkwd{legend}\hlstd{(}\hlstr{"bottomright"}\hlstd{,} \hlkwc{pch}\hlstd{=}\hlstr{"x"}\hlstd{,} \hlkwc{col}\hlstd{=}\hlkwd{c}\hlstd{(}\hlstr{"blue"}\hlstd{,} \hlstr{"red"}\hlstd{),}
       \hlkwc{legend} \hlstd{=} \hlkwd{c}\hlstd{(}\hlstr{"flexmix"}\hlstd{,} \hlstr{"mixdistreg"}\hlstd{))}
\end{alltt}
\end{kframe}
\end{knitrout}
\noindent The results of the above code are shown in Figure~\ref{fig:1}.

\subsection{Mixture of Different Regressions}

The previous data could alternatively also be fitted with a mixture of different distributions. We here choose the normal and Laplace distribution.

\begin{figure}[!h]
    \centering
    \includegraphics[width=0.8\maxwidth]{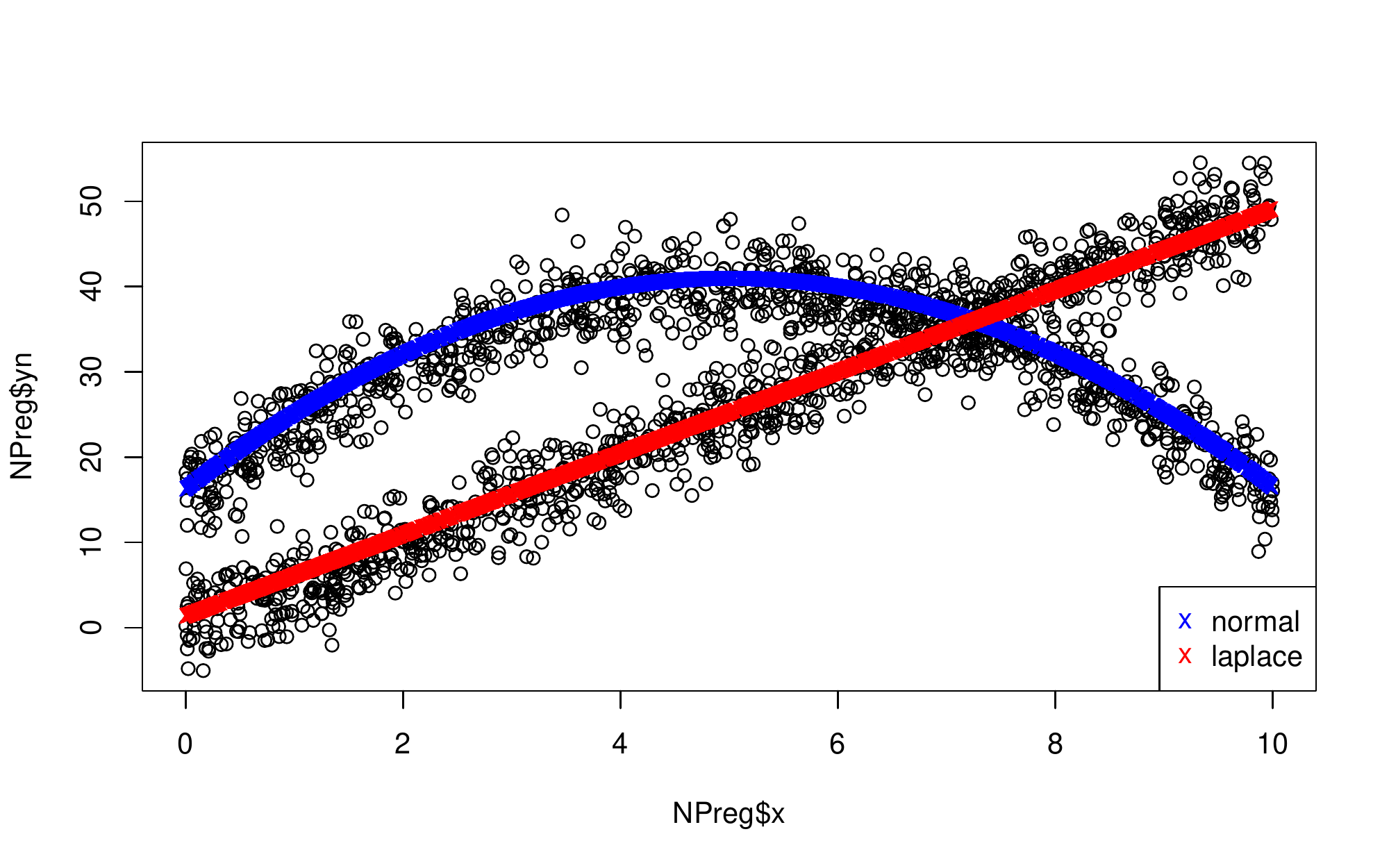}
    \caption{The estimated means of the two different mixture components (normal and laplace distribution).}
    \label{fig:2}
\end{figure}

\begin{knitrout}
\definecolor{shadecolor}{rgb}{0.969, 0.969, 0.969}\color{fgcolor}\begin{kframe}
\begin{alltt}
\hlcom{# Define a different distribution}
\hlstd{dr_mod2} \hlkwb{<-} \hlkwd{mixdistreg}\hlstd{(}\hlkwc{y} \hlstd{= NPreg}\hlopt{$}\hlstd{yn} \hlopt{+} \hlnum{1}\hlstd{,}
                      \hlkwc{families} \hlstd{=} \hlkwd{c}\hlstd{(}\hlstr{"normal"}\hlstd{,} \hlstr{"laplace"}\hlstd{),}
                      \hlkwc{nr_comps} \hlstd{= nr_comps,}
                      \hlkwc{list_of_formulas} \hlstd{=} \hlkwd{list}\hlstd{(}
                        \hlcom{# parameters for normal}
                        \hlkwc{mean} \hlstd{=} \hlopt{~} \hlnum{1} \hlopt{+} \hlstd{x} \hlopt{+} \hlstd{xsq,}
                        \hlkwc{scale} \hlstd{=} \hlopt{~}\hlnum{1}\hlstd{,}
                        \hlcom{# parameters for laplace}
                        \hlkwc{location} \hlstd{=} \hlopt{~} \hlnum{1} \hlopt{+} \hlstd{x} \hlopt{+} \hlstd{xsq,}
                        \hlkwc{scale} \hlstd{=} \hlopt{~}\hlnum{1}
                      \hlstd{),}
                      \hlkwc{data} \hlstd{= NPreg,}
                      \hlkwc{optimizer} \hlstd{=} \hlkwd{optimizer_rmsprop}\hlstd{(}
                        \hlkwc{learning_rate} \hlstd{=} \hlnum{0.01}\hlstd{),}
                      \hlkwc{tf_seed} \hlstd{=} \hlnum{42}
\hlstd{)}

\hlcom{# Train network}
\hlstd{dr_mod2} \hlopt{%>%} \hlkwd{fit}\hlstd{(}\hlkwc{epochs} \hlstd{=} \hlnum{5000L}\hlstd{,}
                \hlkwc{validation_split} \hlstd{=} \hlnum{0.1}\hlstd{,}
                \hlkwc{patience} \hlstd{=} \hlnum{100L}\hlstd{,}
                \hlkwc{early_stopping} \hlstd{=} \hlnum{TRUE}\hlstd{,}
                \hlkwc{verbose} \hlstd{=} \hlnum{FALSE}\hlstd{)}

\hlcom{# Get estimated means distributions}
\hlstd{pred_dr2} \hlkwb{<-} \hlstd{dr_mod2} \hlopt{%>%} \hlkwd{get_stats_mixcomps}\hlstd{(}\hlkwc{what} \hlstd{=} \hlstr{"means"}\hlstd{)}

\hlkwd{plot}\hlstd{(NPreg}\hlopt{$}\hlstd{yn} \hlopt{~} \hlstd{NPreg}\hlopt{$}\hlstd{x)}
\hlkwa{for}\hlstd{(i} \hlkwa{in} \hlnum{1}\hlopt{:}\hlstd{nr_comps)\{}
  \hlkwd{points}\hlstd{(pred_dr2[,i]} \hlopt{~} \hlstd{NPreg}\hlopt{$}\hlstd{x,} \hlkwc{col}\hlstd{=}\hlkwd{c}\hlstd{(}\hlstr{"blue"}\hlstd{,}\hlstr{"red"}\hlstd{)[i],} \hlkwc{pch}\hlstd{=}\hlstr{"x"}\hlstd{)}
\hlstd{\}}
\hlkwd{legend}\hlstd{(}\hlstr{"bottomright"}\hlstd{,} \hlkwc{pch}\hlstd{=}\hlstr{"x"}\hlstd{,} \hlkwc{col}\hlstd{=}\hlkwd{c}\hlstd{(}\hlstr{"blue"}\hlstd{,} \hlstr{"red"}\hlstd{),}
       \hlkwc{legend} \hlstd{=} \hlkwd{c}\hlstd{(}\hlstr{"normal"}\hlstd{,} \hlstr{"laplace"}\hlstd{))}
\end{alltt}
\end{kframe}
\end{knitrout}
\noindent The results of the above code are shown in Figure~\ref{fig:2}.

\subsection{Mixture of Experts Distributional Regression}

Next, we change the above data-generation process to include a heterogeneous variance different in both clusters. We then adapt the model to a mixture of experts distributional regression.

\begin{knitrout}
\definecolor{shadecolor}{rgb}{0.969, 0.969, 0.969}\color{fgcolor}\begin{kframe}
\begin{alltt}
\hlcom{# Generate response differently}
\hlkwd{set.seed}\hlstd{(}\hlnum{32}\hlstd{)}
\hlstd{n} \hlkwb{<-} \hlnum{1000}
\hlstd{NPreg}\hlopt{$}\hlstd{yn} \hlkwb{<-}
  \hlkwd{c}\hlstd{(}\hlnum{5} \hlopt{*} \hlstd{NPreg}\hlopt{$}\hlstd{x[}\hlnum{1}\hlopt{:}\hlstd{n]} \hlopt{+} \hlnum{3} \hlopt{*} \hlkwd{rnorm}\hlstd{(n,} \hlnum{0}\hlstd{,} \hlkwd{exp}\hlstd{(}\hlopt{-}\hlnum{1}\hlopt{+}\hlstd{NPreg}\hlopt{$}\hlstd{x}\hlopt{/}\hlnum{5}\hlstd{)),}
    \hlnum{40} \hlopt{-} \hlstd{(NPreg}\hlopt{$}\hlstd{x[(n} \hlopt{+} \hlnum{1}\hlstd{)}\hlopt{:}\hlstd{(}\hlnum{2} \hlopt{*} \hlstd{n)]} \hlopt{-} \hlnum{5}\hlstd{)}\hlopt{^}\hlnum{2} \hlopt{+} \hlnum{3} \hlopt{*} \hlkwd{rnorm}\hlstd{(n))}

\hlcom{# Define a mixture of distributional regressions}
\hlstd{dr_mod3} \hlkwb{<-} \hlkwd{sammer}\hlstd{(}\hlkwc{y} \hlstd{= NPreg}\hlopt{$}\hlstd{yn,}
                  \hlkwc{family} \hlstd{=} \hlstr{"normal"}\hlstd{,}
                  \hlkwc{nr_comps} \hlstd{= nr_comps,}
                  \hlkwc{list_of_formulas} \hlstd{=} \hlkwd{list}\hlstd{(}\hlkwc{mean} \hlstd{=} \hlopt{~} \hlnum{1} \hlopt{+} \hlstd{x} \hlopt{+} \hlstd{xsq,}
                                          \hlkwc{scale} \hlstd{=} \hlopt{~}\hlnum{1} \hlopt{+} \hlstd{x),}
                  \hlkwc{data} \hlstd{= NPreg,}
                  \hlkwc{optimizer} \hlstd{=} \hlkwd{optimizer_rmsprop}\hlstd{(}\hlkwc{learning_rate} \hlstd{=} \hlnum{0.01}\hlstd{),}
                  \hlkwc{tf_seed} \hlstd{=} \hlnum{42}
\hlstd{)}

\hlcom{# Train network}
\hlstd{dr_mod3} \hlopt{%>%} \hlkwd{fit}\hlstd{(}\hlkwc{epochs} \hlstd{=} \hlnum{5000L}\hlstd{,}
                \hlkwc{validation_split} \hlstd{=} \hlnum{0.1}\hlstd{,}
                \hlkwc{patience} \hlstd{=} \hlnum{100L}\hlstd{,}
                \hlkwc{early_stopping} \hlstd{=} \hlnum{TRUE}\hlstd{,}
                \hlkwc{verbose} \hlstd{=} \hlnum{FALSE}\hlstd{)}

\hlcom{# Get estimated mean and standard deviations}
\hlstd{pred_dr3} \hlkwb{<-} \hlstd{dr_mod3} \hlopt{%>%} \hlkwd{get_stats_mixcomps}\hlstd{(}\hlkwc{what} \hlstd{=} \hlstr{"means"}\hlstd{)}
\hlstd{stddev_dr3} \hlkwb{<-} \hlstd{dr_mod3} \hlopt{%>%} \hlkwd{get_stats_mixcomps}\hlstd{(}\hlkwc{what} \hlstd{=} \hlstr{"stddev"}\hlstd{)}

\hlkwd{plot}\hlstd{(NPreg}\hlopt{$}\hlstd{yn} \hlopt{~} \hlstd{NPreg}\hlopt{$}\hlstd{x)}
\hlkwa{for}\hlstd{(i} \hlkwa{in} \hlnum{1}\hlopt{:}\hlstd{nr_comps)\{}
  \hlkwd{points}\hlstd{(pred_dr3[,i]}\hlopt{~}\hlstd{NPreg}\hlopt{$}\hlstd{x,} \hlkwc{col}\hlstd{=}\hlkwd{c}\hlstd{(}\hlstr{"blue"}\hlstd{,}\hlstr{"red"}\hlstd{)[i],} \hlkwc{pch}\hlstd{=}\hlstr{"x"}\hlstd{)}
  \hlkwd{points}\hlstd{(pred_dr3[,i]}\hlopt{+}\hlnum{2}\hlopt{*}\hlstd{stddev_dr3[,i]}\hlopt{~}\hlstd{NPreg}\hlopt{$}\hlstd{x,} \hlkwc{col}\hlstd{=}\hlkwd{c}\hlstd{(}\hlstr{"blue"}\hlstd{,}\hlstr{"red"}\hlstd{)[i],} \hlkwc{pch}\hlstd{=}\hlstr{"-"}\hlstd{)}
  \hlkwd{points}\hlstd{(pred_dr3[,i]}\hlopt{-}\hlnum{2}\hlopt{*}\hlstd{stddev_dr3[,i]}\hlopt{~}\hlstd{NPreg}\hlopt{$}\hlstd{x,} \hlkwc{col}\hlstd{=}\hlkwd{c}\hlstd{(}\hlstr{"blue"}\hlstd{,}\hlstr{"red"}\hlstd{)[i],} \hlkwc{pch}\hlstd{=}\hlstr{"-"}\hlstd{)}
\hlstd{\}}
\hlkwd{legend}\hlstd{(}\hlstr{"bottomright"}\hlstd{,} \hlkwc{pch}\hlstd{=}\hlstr{"x"}\hlstd{,} \hlkwc{col}\hlstd{=}\hlkwd{c}\hlstd{(}\hlstr{"blue"}\hlstd{,} \hlstr{"red"}\hlstd{),}
       \hlkwc{legend} \hlstd{=} \hlkwd{c}\hlstd{(}\hlstr{"Comp. 1"}\hlstd{,} \hlstr{"Comp. 2"}\hlstd{))}
\end{alltt}
\end{kframe}
\end{knitrout}

\begin{figure}[!h]
    \centering
    \includegraphics[width=0.8\maxwidth]{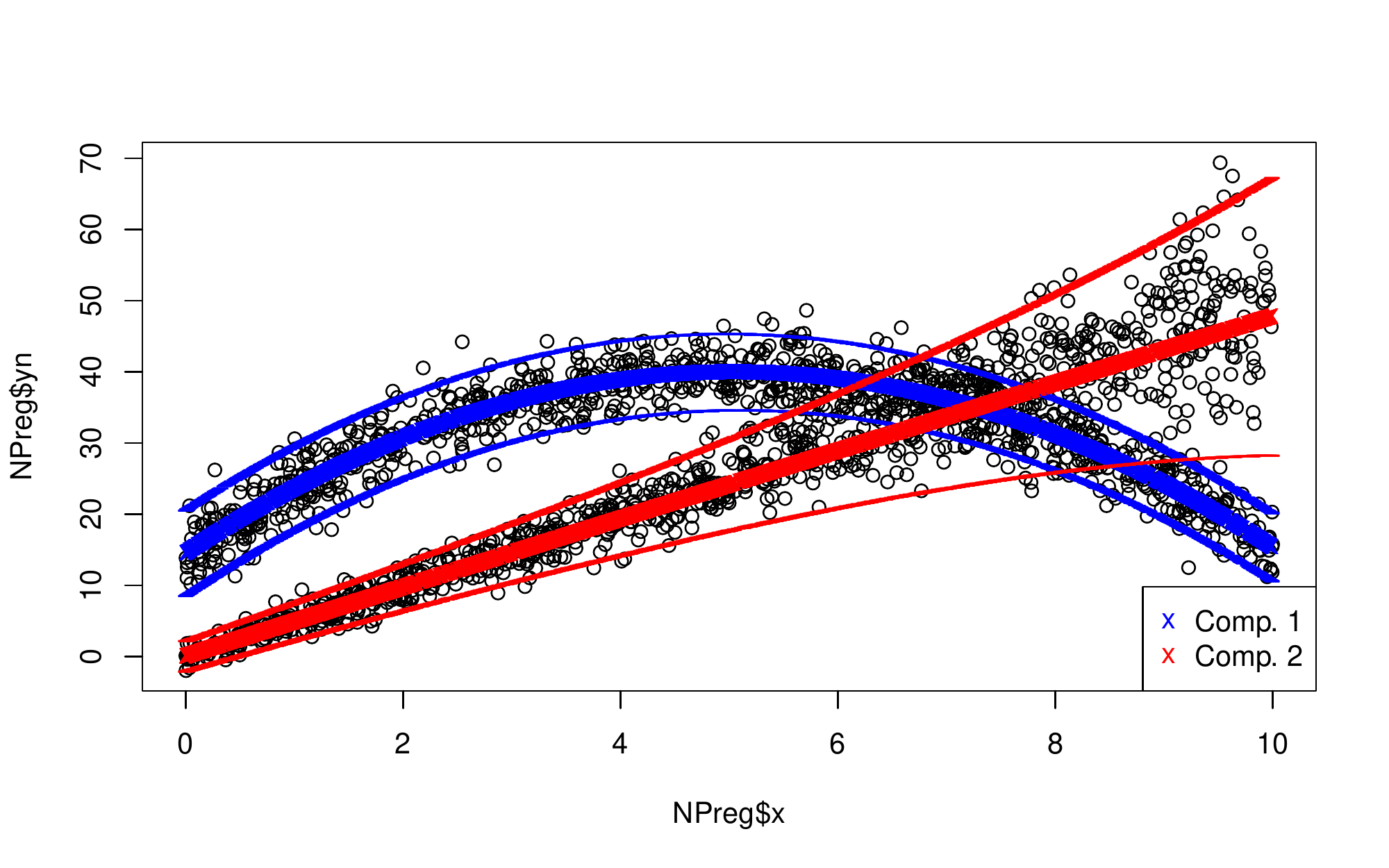}
    \caption{The estimated means and $\pm$ two times the estimated standard deviations of the two normal distributions fitted on a heterogeneous version of the \texttt{NPreg} data where one component has an increased variance for larger \texttt{x} values.}
    \label{fig:3}
\end{figure}

\noindent The results of the above code are shown in Figure~\ref{fig:3}.

\subsection{Semi-structured Mixture Density Networks}

In contrast to the previous examples, we now construct a mixture model that learns one of the mixtures using an (unstructured) deep neural network in the style of a mixture density network. The second distribution is again learned with a structured predictor. We use the data from the previous subsection to demonstrate this.

\begin{knitrout}
\definecolor{shadecolor}{rgb}{0.969, 0.969, 0.969}\color{fgcolor}\begin{kframe}
\begin{alltt}
\hlcom{# Deep network}
\hlstd{deep_net} \hlkwb{<-} \hlkwd{keras_model_sequential}\hlstd{()} \hlopt{%>%}
  \hlkwd{layer_dense}\hlstd{(}\hlkwc{units} \hlstd{=} \hlnum{64}\hlstd{,} \hlkwc{activation} \hlstd{=} \hlstr{"relu"}\hlstd{,} \hlkwc{use_bias} \hlstd{=} \hlnum{FALSE}\hlstd{)} \hlopt{%>%}
  \hlkwd{layer_dense}\hlstd{(}\hlkwc{units} \hlstd{=} \hlnum{64}\hlstd{,} \hlkwc{activation} \hlstd{=} \hlstr{"relu"}\hlstd{,} \hlkwc{use_bias} \hlstd{=} \hlnum{FALSE}\hlstd{)} \hlopt{%>%}
  \hlkwd{layer_dense}\hlstd{(}\hlkwc{units} \hlstd{=} \hlnum{32}\hlstd{,} \hlkwc{activation} \hlstd{=} \hlstr{"relu"}\hlstd{,} \hlkwc{use_bias} \hlstd{=} \hlnum{FALSE}\hlstd{)}
\end{alltt}

\begin{alltt}
\hlcom{# Deep network head for the mean}
\hlstd{deep_mean} \hlkwb{<-} \hlkwa{function}\hlstd{(}\hlkwc{x}\hlstd{) x} \hlopt{%>%}
  \hlstd{deep_net} \hlopt{%>%}
  \hlkwd{layer_dense}\hlstd{(}\hlkwc{units} \hlstd{=} \hlnum{1}\hlstd{)}

\hlcom{# Deep network head for the standard deviation}
\hlstd{deep_std} \hlkwb{<-} \hlkwa{function}\hlstd{(}\hlkwc{x}\hlstd{) x} \hlopt{%>%}
  \hlstd{deep_net} \hlopt{%>%}
  \hlkwd{layer_dense}\hlstd{(}\hlkwc{units} \hlstd{=} \hlnum{1}\hlstd{,} \hlkwc{activation} \hlstd{=} \hlstr{"softplus"}\hlstd{)}

\hlcom{# Semi-structured mixture density network}
\hlstd{dr_mod4} \hlkwb{<-} \hlkwd{mixdistreg}\hlstd{(}\hlkwc{y} \hlstd{= NPreg}\hlopt{$}\hlstd{yn,}
                      \hlkwc{families} \hlstd{=} \hlkwd{c}\hlstd{(}\hlstr{"normal"}\hlstd{,} \hlstr{"normal"}\hlstd{),}
                      \hlkwc{nr_comps} \hlstd{= nr_comps,}
                      \hlkwc{list_of_formulas} \hlstd{=} \hlkwd{list}\hlstd{(}
                        \hlcom{# structured model part}
                        \hlkwc{mean1} \hlstd{=} \hlopt{~} \hlnum{1} \hlopt{+} \hlstd{x} \hlopt{+} \hlstd{xsq,}
                        \hlkwc{scale1} \hlstd{=} \hlopt{~}\hlnum{1} \hlopt{+} \hlstd{x,}
                        \hlcom{# unstructured deep network}
                        \hlkwc{mean1} \hlstd{=} \hlopt{~} \hlnum{1} \hlopt{+} \hlkwd{dm}\hlstd{(x),}
                        \hlkwc{scale1} \hlstd{=} \hlopt{~}\hlnum{1} \hlopt{+} \hlkwd{ds}\hlstd{(x)}
                        \hlstd{),}
                      \hlkwc{list_of_deep_models} \hlstd{=} \hlkwd{list}\hlstd{(}\hlkwc{dm} \hlstd{= deep_mean,}
                                                 \hlkwc{ds} \hlstd{= deep_std),}
                      \hlkwc{data} \hlstd{= NPreg,}
                      \hlkwc{optimizer} \hlstd{=} \hlkwd{optimizer_adam}\hlstd{(}\hlkwc{learning_rate} \hlstd{=} \hlnum{0.001}\hlstd{),}
                      \hlkwc{tf_seed} \hlstd{=} \hlnum{42}
\hlstd{)}

\hlcom{# Train network}
\hlstd{dr_mod4} \hlopt{%>%} \hlkwd{fit}\hlstd{(}\hlkwc{epochs} \hlstd{=} \hlnum{5000L}\hlstd{,}
                \hlkwc{validation_split} \hlstd{=} \hlnum{0.1}\hlstd{,}
                \hlkwc{patience} \hlstd{=} \hlnum{500L}\hlstd{,}
                \hlkwc{early_stopping} \hlstd{=} \hlnum{TRUE}\hlstd{,}
                \hlkwc{verbose} \hlstd{=} \hlnum{FALSE}\hlstd{)}

\hlcom{# Extract means and standard deviations}
\hlstd{pred_dr4} \hlkwb{<-} \hlstd{dr_mod4} \hlopt{%>%} \hlkwd{get_stats_mixcomps}\hlstd{(}\hlkwc{what} \hlstd{=} \hlstr{"means"}\hlstd{)}
\hlstd{stddev_dr4} \hlkwb{<-} \hlstd{dr_mod4} \hlopt{%>%} \hlkwd{get_stats_mixcomps}\hlstd{(}\hlkwc{what} \hlstd{=} \hlstr{"stddev"}\hlstd{)}

\hlkwd{plot}\hlstd{(NPreg}\hlopt{$}\hlstd{yn} \hlopt{~} \hlstd{NPreg}\hlopt{$}\hlstd{x)}
\hlkwa{for}\hlstd{(i} \hlkwa{in} \hlnum{1}\hlopt{:}\hlstd{nr_comps)\{}
  \hlkwd{points}\hlstd{(pred_dr4[,i]}\hlopt{~}\hlstd{NPreg}\hlopt{$}\hlstd{x,} \hlkwc{col}\hlstd{=}\hlkwd{c}\hlstd{(}\hlstr{"blue"}\hlstd{,}\hlstr{"red"}\hlstd{)[i],} \hlkwc{pch}\hlstd{=}\hlstr{"x"}\hlstd{)}
  \hlkwd{points}\hlstd{(pred_dr4[,i]}\hlopt{+}\hlnum{2}\hlopt{*}\hlstd{stddev_dr4[,i]}\hlopt{~}\hlstd{NPreg}\hlopt{$}\hlstd{x,} \hlkwc{col}\hlstd{=}\hlkwd{c}\hlstd{(}\hlstr{"blue"}\hlstd{,}\hlstr{"red"}\hlstd{)[i],} \hlkwc{pch}\hlstd{=}\hlstr{"-"}\hlstd{)}
  \hlkwd{points}\hlstd{(pred_dr4[,i]}\hlopt{-}\hlnum{2}\hlopt{*}\hlstd{stddev_dr4[,i]}\hlopt{~}\hlstd{NPreg}\hlopt{$}\hlstd{x,} \hlkwc{col}\hlstd{=}\hlkwd{c}\hlstd{(}\hlstr{"blue"}\hlstd{,}\hlstr{"red"}\hlstd{)[i],} \hlkwc{pch}\hlstd{=}\hlstr{"-"}\hlstd{)}
\hlstd{\}}
\hlkwd{legend}\hlstd{(}\hlstr{"bottomright"}\hlstd{,} \hlkwc{pch}\hlstd{=}\hlstr{"x"}\hlstd{,} \hlkwc{col}\hlstd{=}\hlkwd{c}\hlstd{(}\hlstr{"blue"}\hlstd{,} \hlstr{"red"}\hlstd{),}
       \hlkwc{legend} \hlstd{=} \hlkwd{c}\hlstd{(}\hlstr{"Comp. 1"}\hlstd{,} \hlstr{"Comp. 2"}\hlstd{))}
\end{alltt}
\end{kframe}
\end{knitrout}

\begin{figure}[!h]
    \centering
    \includegraphics[width=0.8\maxwidth]{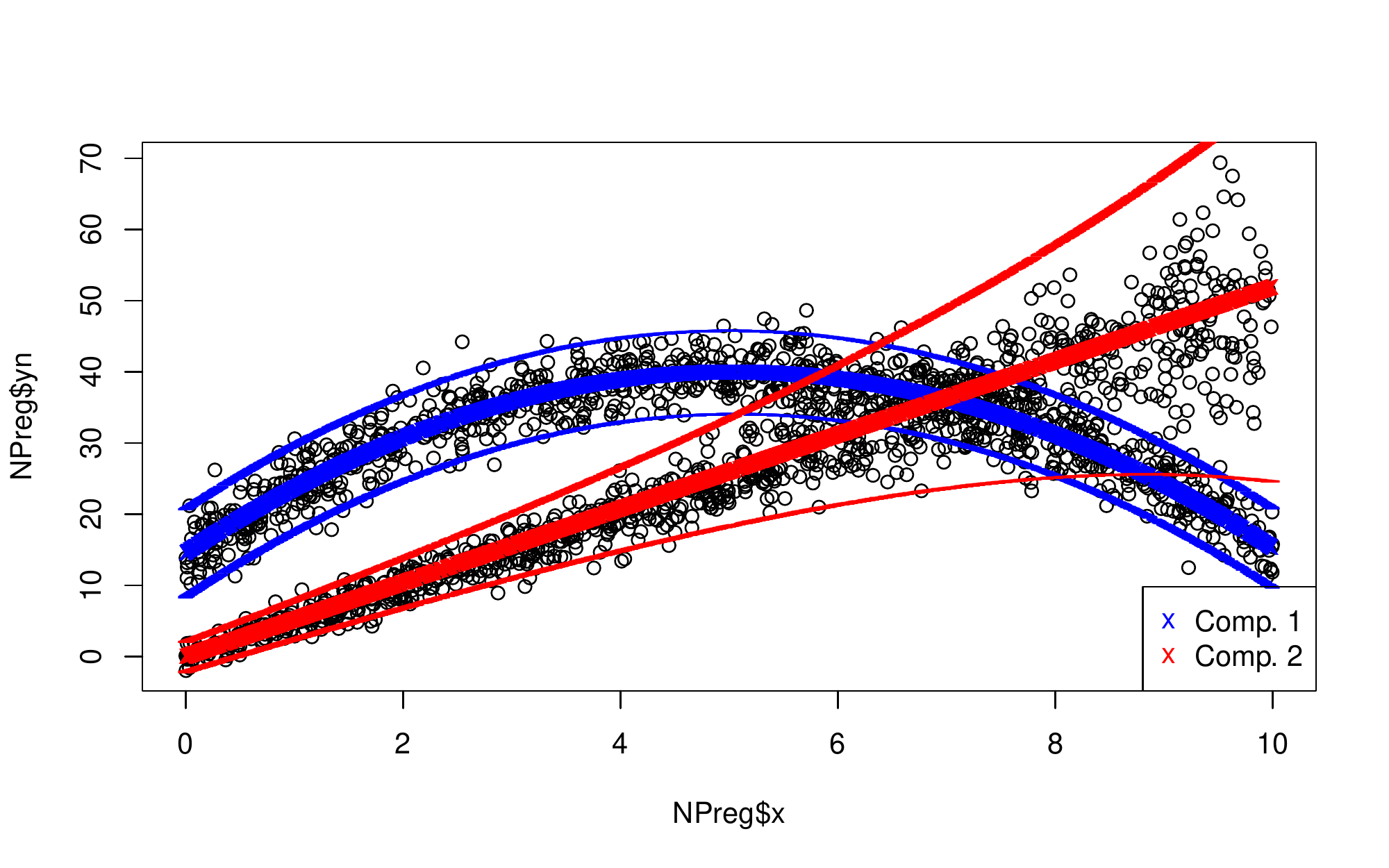}
    \caption{The estimated means and $\pm$ two times the estimated standard deviations of the two normal distributions fitted on a heterogeneous version of the \texttt{NPreg} data where one component is learned using a deep neural network.}
    \label{fig:4}
\end{figure}

\noindent The results of the above code are shown in Figure~\ref{fig:4}.

\subsection{Zero-and-one Inflated Regression}

As a last example, we generate zero-inflated data and show how to use the package's wrapper function \texttt{zinreg} to estimate a mixture of a point distribution at zero and a normal distribution.

\begin{knitrout}
\definecolor{shadecolor}{rgb}{0.969, 0.969, 0.969}\color{fgcolor}\begin{kframe}
\begin{alltt}
\hlcom{# zero-inflated}
\hlkwd{set.seed}\hlstd{(}\hlnum{32}\hlstd{)}
\hlstd{n} \hlkwb{<-} \hlnum{1000}
\hlstd{NPreg}\hlopt{$}\hlstd{yn} \hlkwb{<-}
  \hlkwd{c}\hlstd{(}\hlnum{5} \hlopt{*} \hlstd{NPreg}\hlopt{$}\hlstd{x[}\hlnum{1}\hlopt{:}\hlstd{n]} \hlopt{+} \hlnum{3} \hlopt{*} \hlkwd{rnorm}\hlstd{(n,} \hlnum{0}\hlstd{,} \hlkwd{exp}\hlstd{(}\hlopt{-}\hlnum{1}\hlopt{+}\hlstd{NPreg}\hlopt{$}\hlstd{x}\hlopt{/}\hlnum{5}\hlstd{)),}
    \hlkwd{rep}\hlstd{(}\hlnum{0}\hlstd{, n))}

\hlcom{# Zero-inflated regression with normal distribution}
\hlstd{dr_mod4} \hlkwb{<-} \hlkwd{zinreg}\hlstd{(}\hlkwc{y} \hlstd{= NPreg}\hlopt{$}\hlstd{yn,}
                  \hlkwc{family} \hlstd{=} \hlstr{"normal"}\hlstd{,}
                  \hlkwc{list_of_formulas} \hlstd{=} \hlkwd{list}\hlstd{(}\hlkwc{mean} \hlstd{=} \hlopt{~} \hlnum{1} \hlopt{+} \hlstd{x} \hlopt{+} \hlstd{xsq,}
                                          \hlkwc{scale} \hlstd{=} \hlopt{~}\hlnum{1} \hlopt{+} \hlstd{x),}
                  \hlkwc{data} \hlstd{= NPreg,}
                  \hlkwc{optimizer} \hlstd{=} \hlkwd{optimizer_rmsprop}\hlstd{(}\hlkwc{learning_rate} \hlstd{=} \hlnum{0.01}\hlstd{),}
                  \hlkwc{tf_seed} \hlstd{=} \hlnum{42}
\hlstd{)}

\hlcom{# Train network}
\hlstd{dr_mod4} \hlopt{%>%} \hlkwd{fit}\hlstd{(}\hlkwc{epochs} \hlstd{=} \hlnum{5000L}\hlstd{,}
                \hlkwc{validation_split} \hlstd{=} \hlnum{0.1}\hlstd{,}
                \hlkwc{patience} \hlstd{=} \hlnum{100L}\hlstd{,}
                \hlkwc{early_stopping} \hlstd{=} \hlnum{TRUE}\hlstd{,}
                \hlkwc{verbose} \hlstd{=} \hlnum{FALSE}\hlstd{)}

\hlcom{# Check estimated probabilities}
\hlstd{(dr_mod4} \hlopt{%>%} \hlkwd{get_pis}\hlstd{())[}\hlnum{1}\hlstd{,]}
\end{alltt}
\begin{verbatim}
## [1] 0.5379843 0.4620157
\end{verbatim}
\end{kframe}
\end{knitrout}

\noindent Results show that the probability for zero-inflation is learned almost correctly. Using the function $\texttt{inflareg}$, the previous code could be adapted to arbitrary value-inflated distributions, also for more than one value (e.g., for zero-one-inflation).


\bibliographystyle{splncs04}
\setlength{\bibsep}{0pt plus 0.3ex}
\begin{footnotesize}
\bibliography{references}
\end{footnotesize}

\end{document}